\documentclass[conference]{IEEEtran}
\usepackage[mathcal]{euscript}
\usepackage{amsmath}
\usepackage{amsthm}
\usepackage{amssymb}
\usepackage{graphicx}
\usepackage{algorithmic}
\usepackage{algorithm}
\usepackage{epsfig}
\usepackage{epstopdf}
\usepackage{booktabs}
\usepackage{colortbl}
\usepackage{amsmath}
\usepackage{xcolor}
\usepackage{graphicx}
\usepackage{cite} 

\usepackage{authblk}

\begin{document}

\title{Supervised Machine Learning for Signals Having RRC Shaped Pulses}

\author[ 1]{Mohammad Bari}
\author[ 2]{Hussain Taher}
\author[ 2]{Syed Saad Sherazi}
\author[ 1]{Milo\v s Doroslova\v cki}
\affil[ 1]{Electrical and Computer Engineering, The George Washington University, Washington, DC, USA}
\affil[ 2]{Electrical Engineering, University of Engineering \& Technology Peshawar, Pakistan}


\maketitle

\begin{abstract}

Classification performances of the supervised machine learning techniques such as support vector machines, neural networks and logistic regression are compared for modulation recognition purposes. The simple and robust features are used to distinguish continuous-phase FSK from QAM-PSK signals. Signals having root-raised-cosine shaped pulses are simulated in extreme noisy conditions having joint impurities of block fading, lack of symbol and sampling synchronization, carrier offset, and additive white Gaussian noise. The features are based on sample mean and sample variance of the imaginary part of the product of two consecutive complex signal values.  
\end{abstract}


%
\IEEEpeerreviewmaketitle

\section{Introduction}
\label{int}

Signal separation (SS) is used in applications such as interference identification, electronic warfare, enforcement of civilian spectrum compliance, radar, intelligent modems, cognitive and software defined radios. Most of the published work involving the above mentioned communication systems assume that the signals and their parameters are known. For instance , the works in \cite{ewaisha13} and \cite{ewaisha15} do not focus on the SS part. Understandably the focus is on the application itself. This paper studies the SS in noisy conditions that serve as a prerequisite for several applications including but not limited to those mentioned above.

The features in this work, discussed for several scenarios in \cite{Bari15spl}\nocite{Bari15asilomar1}\nocite{Bari14asilomar}\nocite{Bari15cssp}-\cite{Bari13asilomar}, separate the continuous phase frequency shift keying (CPFSK) modulation from linear modulations of phase shift keying (PSK) and quadrature amplitude modulation (QAM). Root raised cosine (RRC) shaped pulses are used to generate the signals. Modulations are simulated to have the joint presence of block fading, lack of symbol and sampling synchronization, carrier offset, additive white Gaussian noise (AWGN) and unknown pulse shape (unknown roll-off). None of the mentioned works discuss the effects of block fading on the performance of SS. Also, support vector machines (SVM) are the only pattern recognition-based approach being employed in the above mentioned works. In this paper, block fading is introduced to make the conditions more challenging. Furthermore, this work compares the classification performances of the SVM, logistic regression (LR) and neural networks (NN) for the same set of features. The order of modulation of the CPFSK signals (binary, 4-ary and 8-ary CPFSK) is identified in \cite{Bari15asilomar2}. In \cite{Bari15asilomar3}, frequency modulated signals are separated from the linearly modulated signals.

\section{Signal Model}
\label{sig}

We model the complex baseband continuous-time received signal as \cite{Bari15asilomar1}
\begin{equation}
s(t)=
x(t-t_0)e^{j(\Delta t+\theta_c)}\alpha(t)e^{j\psi(t)}+v(t)
\label{e1}
\end{equation}
where $x(t)$ is the transmitted-signal, $\theta_c$ is the initial phase
uniformly distributed over $[0,~2\pi)$, $\Delta$ is the
carrier offset, $\alpha(t)e^{j\psi(t)}$ models the fading. $\alpha(t)$ is a Rayleigh random
variable. $\psi(t)$ is uniformly distributed over $[0,2\pi)$ and is independent of $\alpha(t)$. $v(t)$ is zero-mean complex noise and $t_0$ is the time delay. Furthermore, it is assumed that the bandwidth $B$ of the receiver's filter is in general
larger than the bandwidth of the transmitted signal and the
power spectral density of $v(t)$ does not vary within the filter's
bandwidth. In this work the center of the signal's spectrum is translated to be around a certain desired normalized frequency, say $0$ or $\pi/2$. 

For linear modulations the transmitted-signal is
\begin{equation}
x(t-t_0)=
\sum\limits_{n=-\infty}^{+\infty}a_ne^{j\theta_n}p(t-t_0-nT)
\label{e2}
\end{equation}
where $(a_n,~\theta_n)$ are the amplitude and phase of the transmitted symbol, $p(t)$ is the pulse shape function and $T$ is the symbol period. In the case of FSK modulation
\begin{equation}
x(t-t_0)=
e^{j\int\limits_{0}^{t-t_0}\sum\limits_{n=-\infty}^{+\infty}b_nq(\rho-nT)d\rho}
\label{e3}
\end{equation}
where $q(\rho)$ defines instantaneous frequency pulse shape and $b_n\in\{-1,+1\}$ for BFSK.
Note that (\ref{e3}) models continuous phase FSK, which is of higher practical interest than non-continuous phase FSK used in \cite{Bari13ciss}. 

Sampling period, $T_s$, is assumed to be $1/(2B)$. Symbol period is given by 
\begin{equation}
T=N_sT_s+\varepsilon T_s
\label{eT}
\end{equation}
where $N_s$ is the oversampling and $\varepsilon$ is uniformly distributed in $[0,1)$.
Finally the complex baseband discrete-time received signal is 
\begin{equation}
s[k]\hspace{-1mm}=\hspace{-1mm}
x(kT_s-t_0)e^{j(\Delta' k+\theta_c)}\alpha[k]e^{j\psi[k]}\hspace{-1mm}+v[k]\hspace{-1mm}=\hspace{-1mm}s(t)\vert_{t=kT_s}\label{e13}
\end{equation}
where $\Delta'=\Delta T_s$, $\alpha[k]=\alpha(kT_s)$, $\psi[k]=\psi(kT_s)$ and $v[k]$ is complex circular AWGN having zero mean and unit variance. Note that $t_0=k_0T_s+\varepsilon_0T_s$ where $k_0$ is the integer part and $\varepsilon_0\in[0,1)$ is the
fractional part of the time delay $t_0$ measured in sampling
periods as time units.

\section{The Three Features}
\label{der}

Let 
\begin{equation}
w[k]=s[k]s^*[k-1]
\label{e14}
\end{equation}
where * is the complex conjugate operator. We assume that $\alpha[k]=1,~\psi[k]=0$, i.e., fading is skipped for the sake of simplicity. $w[k]$ for noiseless QAM-PSK signals is
\begin{equation}
\begin{split}
\hspace{-2mm}
w[k]\vert_{v[k]\equiv 0} &=\hspace{-2mm}
\sum\limits_{n=-\infty}^{+\infty}a_n^2e^{j\Delta'}P_n[k]P_n^-[k]+
\\&~~~\hspace{-5mm}
\sum\limits_{n=-\infty}^{+\infty}\sum\limits_{m=-\infty}^{+\infty}a_na_me^{j[\theta_n-\theta_m+\Delta']}P_n[k]P_m^-[k]
\end{split}
\label{e15}
\end{equation}
where $P_n[k]P_m^-[k]=p(kT_s-t_0-nT)p(kT_s-T_s-t_0-mT).$
Similarly, applying $w[k]$ on noiseless FSK signals yields 
\begin{equation}
\begin{split}
w[k]\vert_{v[k]\equiv 0}&=
e^{j[\int\limits_{kT_s-T_s-t_0}^{kT_s-t_0}\omega'(\rho)\frac{d\rho}{T_s}+\Delta']}.
\end{split}
\label{e16}
\end{equation}

Let us consider the means of imaginary part of $w[k]$ in (\ref{e15}) and (\ref{e16}) for linear and BFSK modulations, respectively. For equiprobable constellation points of each modulation, mean of imaginary part of $w[k]$ for 16-QAM, BPSK, 4-PSK and 8-PSK signals is given by \vspace{-1mm}
\begin{equation}
\begin{split}
\hspace{-4mm}E[\mbox{Im}(w[k]\vert_{v[k]\equiv 0})]=&E[\mbox{Im}(w[k])]
\\&\hspace{-4mm}
=\mbox{sin}(\Delta')\sum\limits_{n=-\infty}^{+\infty}P_n[k]P_n^-[k]E[a_n^2]
\label{e17}
\end{split}
\end{equation}
where 16-QAM has constellation points $a_ne^{j\theta_n}\in\{k/\sqrt{10}+jl/\sqrt{10};~k,~l=-3,-1,+1,+3\}$, BPSK's phases $\theta_n\in\{0,~\pi\}$, 4-PSK's phases $\theta_n\in\{(2n+1)\pi/4;~n=0,1,2,3\}$, and 8-PSK's phases $\theta_n\in\{(2n+1)\pi/8;~n=0,1,...,7\}$. Constellation points for 16-QAM, BPSK, 4-PSK and 8-PSK are chosen such that the average power is unity. Mean of imaginary part of $w[k]$ in (\ref{e16}) for BFSK signal is 
\begin{equation}
E[\mbox{Im}(w[k]\vert_{v[k]\equiv 0})]\hspace{-1mm}=\hspace{-1mm}E[\mbox{Im}(w[k])]\hspace{-1mm}=\hspace{-1mm}\sin(\Delta')\hspace{-3mm}\prod_{m=-\infty}^{\infty}\hspace{-3mm}\cos(Q_m[k])
\label{e18}
\end{equation}
where 
\begin{equation}
Q_m[k]=\int\limits_{kT_s-T_s-t_0}^{kT_s-t_0}q(p-mT)d\rho.
\label{e16a}
\end{equation}

Next, let us consider the variances of imaginary part of $w[k]$ in (\ref{e15}) for QAM and PSK modulations, and in (\ref{e16}) for BFSK modulation. The variance for BPSK is $\vspace{-2mm}$
\begin{equation}
\begin{split}
\hspace{-2mm}\mbox{VAR}[\mbox{Im}(w[k]\vert_{v[k]\equiv 0})]&\hspace{-1mm}=\hspace{-1mm}
\sin^2(\Delta')\big[(\hspace{-2mm}\sum\limits_{m=-\infty}^{+\infty}\hspace{-3mm}P_m[k]P_m^-[k])^2+
\\&
\hspace{-18mm}\hspace{-3mm}\sum\limits_{m=-\infty}^{+\infty}\hspace{-3mm}P_m^2[k]\hspace{-3mm}\sum\limits_{m=-\infty}^{+\infty}\hspace{-3mm}(P_m^-[k])^2\hspace{-1mm}-\hspace{-1mm}2\hspace{-3mm}\sum\limits_{m=-\infty}^{+\infty}\hspace{-3mm}(P_m[k]P_m^-[k])^2\big].
\end{split}
\label{e18b}
\end{equation}
For QAM, 4-PSK and 8-PSK modulations, whose signal constellations are invariant to $\pi/2$ rotation, the variance is $\vspace{-2mm}$
\begin{equation}
\begin{split}
\mbox{VAR}[\mbox{Im}(w[k]\vert_{v[k]\equiv 0})]\hspace{-1mm}&=\hspace{-1mm}
\sin^2(\Delta')\bigg[(E[a_0^4]\hspace{-1mm}-\hspace{-1mm}2E^2[a_0^2])\hspace{-3mm}
\\&\hspace{-31mm}
\sum\limits_{m=-\infty}^{+\infty}\hspace{-3mm}(P_m[k]P_m^-[k])^2
\hspace{-1mm}+\hspace{-1mm}E^2[a_0^2](\hspace{-2mm}\sum\limits_{m=-\infty}^{+\infty}\hspace{-3mm}P_m[k]P_m^-[k])^2\hspace{-0.5mm}\bigg]\hspace{-1.5mm}
+\hspace{-1.1mm}\frac{1}{2}E^2[a_0^2]\times
\\&
\hspace{-29mm}
\bigg[\sum\limits_{m=-\infty}^{+\infty}P_m^2[k]\sum\limits_{m=-\infty}^{+\infty}(P_m^-[k])^2-(\sum\limits_{m=-\infty}^{+\infty}P_m[k]P_m^-[k])^2\bigg].
\end{split}
\label{e18a}
\end{equation}
The variance of $\mbox{Im}(w[k])$ in (\ref{e16}) for BFSK modulation is
\begin{equation}
\begin{split}
\hspace{-2mm}\mbox{VAR}[\mbox{Im}(w[k]\vert_{v[k]\equiv 0})]\hspace{-1mm}&=\hspace{-1mm}\frac{1}{2}\hspace{-1mm}-\hspace{-1mm}\frac{1}{2}\hspace{-2mm}\prod\limits_{m=-\infty}^{+\infty}\hspace{-2mm}\cos(2Q_m[k])+
\\&
\hspace{-25mm}\sin^2(\Delta')\big(\hspace{-2mm}\prod\limits_{m=-\infty}^{+\infty}\hspace{-2mm}\cos(2Q_m[k])\hspace{-1mm}-\hspace{-2mm}\prod\limits_{m=-\infty}^{+\infty}\hspace{-2mm}\cos^2(Q_m[k])\big).
\end{split}
\label{e18c}
\end{equation}

\section{Supervised Learning Techniques}
\label{sl}

In this work, classification performances of the SVM, LR and NN are compared for the features introduced in \cite{Bari15spl}. The features are 
\begin{enumerate}
\item  sample mean of $\mbox{Im}(w[k])$ for the signal $s[k]$ obtained by frequency downconversion to $\pi/2$, 
\item sample variance of $\mbox{Im}(w[k])$ for the signal $s[k]$ obtained by frequency downconversion to 0, and 
\item sample variance of $\mbox{Im}(w[k])$  for the signal $s[k]$ obtained by frequency downconversion to $\pi/2$. 
\end{enumerate}
The features based on $\mbox{Im}(w[k])$ for the signal $s[k]$ obtained by frequency downconversion to $\pi/2$ can be seen as based on $\mbox{Re}(w[k])$ for the signal $s[k]$ obtained by frequency downconversion to 0.  

\subsection{SVM}
\label{svm}
SVM are initially introduced in \cite{vapnik63}. The current soft margin form of SVM is presented in \cite{cortes95}. As in any pattern recognition-based approach, SVM create a classification model for the features of the known set of training realizations. The classification model can be linear as well as non-linear. This work uses the linear kernel because the number of features is small and a simple linear decision boundary suffices to separate the signals. For higher dimensional feature space, a more complicated decision boundary, a non-linear one, should be used to avoid the underfitting. It is desirable to have the separation between the features of two classes as wide as possible. The features of the new realizations are then predicted to belong to the either class depending upon which side of the decision boundary they lie. A detailed tutorial can be found in \cite{burges98}.

\subsection{LR}
\label{LR}
For $m$ training examples, LR is the problem of maximizing the following log-likelihood
\begin{equation}
L(\theta) = \sum_{i=1}^{m}\big(y_i\mbox{log}(h(x_i)) + (1-y_i)\mbox{log}(1-h(x_i))\big)
\end{equation}
where $x_i$ is the $i^{\mbox{th}}$ training feature vector example. In this work, $x_i$ is a 3-dimensional feature vector $[x_{i1}, x_{i2}, x_{i3}]^T$, 
$y_i$ is the $i^{\mbox{th}}$ target value. $y_i\in\{0,1\}$
and
\begin{equation}
h(x_i) = \frac{1}{1+e^{-\theta_0 - \sum_{j=1}^{u}\theta_{ij}x_{ij}}}
\end{equation}
where $u$ is dimension of the feature vector $x_i$.

\subsection{NN}
\label{NN}
In this work, the input layer has three nodes, that is, the dimension of the feature vector is 3. There is only one hidden layer having 10 nodes. Since this work discusses the binary classification problem, the output layer has 1 node only.

\section{Simulations and Discussion} 
\label{sim}

The simulation experiments are presented illustrating the classification performance of the SVM, LR and NN for the signals having RRC shaped pulses. The performance is measured by the accuracy, which is total number of correct classifications divided by total number of signals. 

For CPFSK signals, the modulation index is defined as
\begin{equation}
h=2f_dT
\label{eh}
\end{equation}
where $f_d$ is the maximum frequency deviation. Using (\ref{eT}) and (\ref{eh}) we get
\begin{equation}
h=(N_s+\epsilon)\delta/\pi
\end{equation}
where $\delta=2\pi f_dT_s$. For BFSK $\delta=\beta'$ where $\beta'/T_s$ is the amplitude of the instantaneous frequency pulse $q(t)$ in (\ref{e3a}). In this work, FSK signals have carrier frequencies $\beta_{mi}'\in\{\Delta'+(2i-(m+1))h\pi/((m-1)(N_s+\epsilon));~m=2,4,8;i=1,2,....,m\}$, where $\Delta'$ is uniformly distributed in $[\gamma'-\pi/20,~\gamma'+\pi/20]$. The value of $h$ is inversely proportional to the channel's spectral efficiency. Therefore CPFSK signals having $h<1$ are more useful than those where $h\ge1$. Also, SS for $h<1$ is the more challenging scenario than that of $h\ge1.$ Because of the presence of block fading, the SS performance for different values of $h<1$ are very similar. Therefore, only the results for $h=1/2$ are presented in this work.

There are 10000 signals for each modulation (BFSK, 4-FSK, 8-FSK, BPSK, 4-PSK, 8-PSK and  16-QAM). Therefore the total number of modulated signals is 70000.  There are 600 symbols in one realization and the oversampling factor $N_s=6$. Carrier offset, $\Delta'$, is uniformly distributed in $[\gamma'-\pi/20$, $\gamma'+\pi/20]$, where $\gamma'\in\{0,~\pi/2\}.$ Roll-off of the RRC pulses is in $\{k/10;~k=1,~2,~3,....,~10\}$. Both $\Delta'$ and roll-off of the RRC pulses are fixed for a realization and they vary independently from realization to realization following uniform distributions. $\varepsilon$ and $\varepsilon_0$ are both uniformly distributed in [0,1), which means that $T$ and time delay are non-integer multiples of sampling period. This results in asynchronicity between sampling instants and symbol period. $\varepsilon$ and $\varepsilon_0$ remain unchanged for a particular realization. $k_0$ is uniformly distributed in $\{0, 1, 2, ..., \lceil N_s+\epsilon\rceil-1\}$. For block fading, $\alpha[k]e^{i\psi[k]}$ is a constant for a realization and varies independently for each realization. $\alpha[k]$ has unit mean square value, that is, $E[\alpha^2[k]]=1$ and $\psi[k]$ is uniformly distributed in $[0,2\pi)$. 

The signals are simulated to have the joint presence of block fading, unknown roll-off, lack of symbol and sampling synchronization, carrier offset and AWGN.

\begin{figure}[t]
\centerline{{\epsfysize=7.25cm\epsfbox{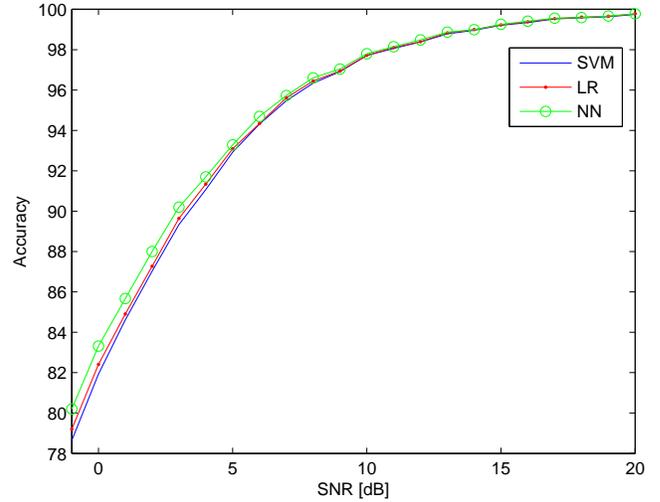}}}
\caption{``SVM'', ``LR'' and ``NN'' represent the performance of the support vector machines, logistic regression and neural network, respectively for training of 1000 realizations for each modulation.}
\label{fig_SLN1000}
\end{figure}

\begin{figure}
\centerline{{\epsfysize=7.25cm\epsfbox{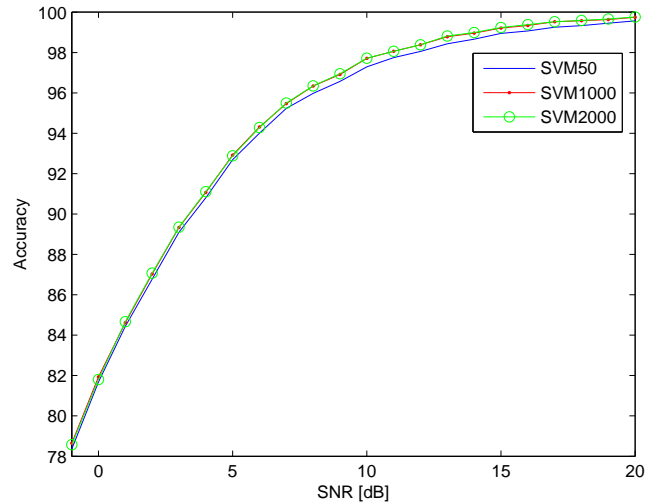}}}
\caption{``SVMx'' represents the performance of the support vector machines for training of $x$ realizations for each modulation.}
\label{fig_SVM}
\end{figure}

\begin{figure}
\centerline{{\epsfysize=7.25cm\epsfbox{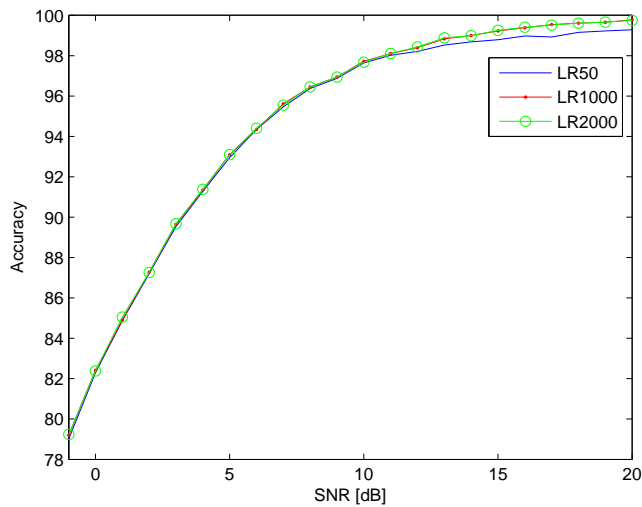}}}
\caption{``LRx'' represents the performance of the logistic regression for training of $x$ realizations for each modulation.}
\label{fig_LR}
\end{figure}

\begin{figure}
\centerline{{\epsfysize=7.25cm\epsfbox{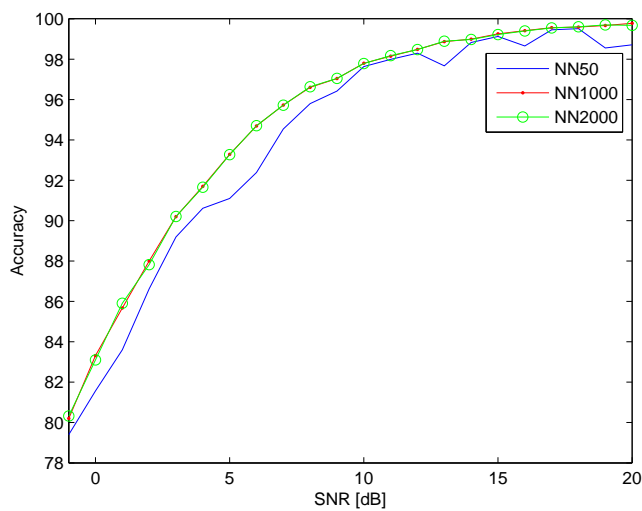}}}
\caption{``NNx'' represents the performance of the neural network for training of $x$ realizations for each modulation.}
\label{fig_NN}
\end{figure}

\begin{figure}
\centerline{{\epsfysize=7.25cm\epsfbox{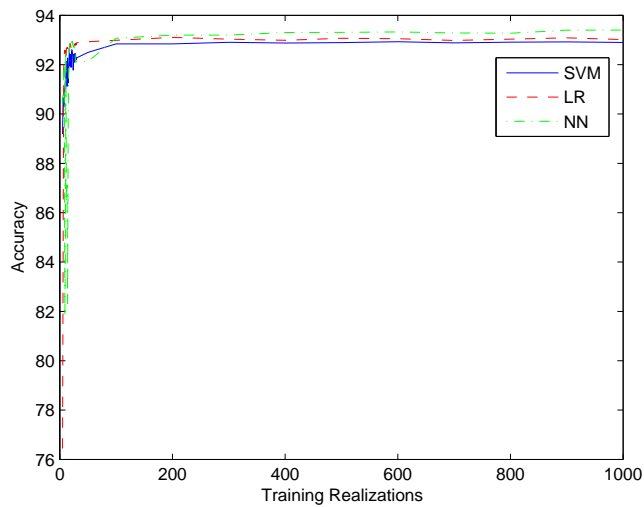}}}
\caption{``SVM'', ``LR'' and ``NN'' represent the performance of the support vector machines, logistic regression and neural network, respectively for SNR = 5 dB.}
\label{fig_SLN1000snrFixed5db}
\end{figure}

Figure \ref{fig_SLN1000} represents the accuracies of the SVM, LR and NN. For each classifier, the number of training realizations is 7000 (1000 realizations for each modulation) for a particular value of SNR. For this large number of training realizations, NN performs the best, especially for smaller values of SNR. LR performs marginally better than the SVM.  

Figures \ref{fig_SVM}, \ref{fig_LR} and \ref{fig_NN} represent the performances of the SVM, LR and NN, respectively. The performances are shown for different number of training realizations. The performance of the particular classifier do not seem to improve when the number of training realizations is increased from 1000 to 2000 for each modulation. Compared to NN, both SVM and LR perform consistently better with the increasing SNRs for the small number of training (50). The accuracies of the SVM and LR for 50 realizations are comparable to those of 1000 and 2000 training realizations. NN, on the other hand, performs poorly for 50 realizations as compared to its performances for 1000 and 2000 training realizations.      

Figure \ref{fig_SLN1000snrFixed5db} represents the accuracies for a range of training realizations for a fixed value of SNR = 5 dB. It can be seen that the NN performs the best for a larger number of training realizations and LR performs marginally better than the SVM.

\section{Conclusions}
\label{con}
The features are classified by SVM, LR and NN, trained for different number of realizations. Each classifier performs similarly for training of 1000 and 2000 realizations. For a low number of 50 training realizations, NN performs poorly. Both SVM and LR perform significantly better than the NN. For a larger number of 1000 or 2000 training realizations, NN performs slightly better than the others for SNRs less than 10 dB. The performances become identical after SNR of 10 dB. If more training realizations are available then NN should be used because of its superior performance for lower values of SNR. If fewer training realizations are available then because of the slightly better performance, especially for lower values of SNR, LR should be preferred over SVM.

\bibliographystyle{IEEEtran}
\bibliography{bib_asilomar2016}
\end{document}